\let\ifarxiv=\iftrue     % ARXIV VERSION
\pdfoutput=1

\ifarxiv

\documentclass[12pt,a4paper]{article}
\usepackage[a4paper,text={450pt,650pt},centering]{geometry}

\fi

\ifarxiv\else

\documentclass[11pt,a4paper]{article}
\usepackage{mathptmx}
\usepackage[a4paper,text={130mm,198mm}]{geometry}

\fi

\usepackage{enumerate}
\usepackage{amsmath,amssymb}
\usepackage[bookmarks=true,hyperfigures=true]{hyperref}
\usepackage{graphicx}
\usepackage[nosort]{cite}
\let\oldbfseries=\bfseries
\let\oldmdseries=\mdseries
\let\oldnormalfont=\normalfont
\renewcommand{\bfseries}{\oldbfseries\boldmath}
\renewcommand{\mdseries}{\oldmdseries\unboldmath}
\renewcommand{\normalfont}{\oldnormalfont\unboldmath}

\allowdisplaybreaks[3]

\numberwithin{equation}{section}

\usepackage[font=small,labelfont=bf,width=0.85\textwidth]{caption}

\providecommand{\hypersetup}[1]{}
\providecommand{\texorpdfstring}[2]{#1}

\hypersetup{plainpages=false}
\hypersetup{pdfpagemode=UseNone}
\hypersetup{bookmarksnumbered=true}
\hypersetup{pdfstartview=FitH}
\hypersetup{colorlinks=false}
\hypersetup{citebordercolor={.5 1 .5}}
\hypersetup{urlbordercolor={.5 1 1}}
\hypersetup{linkbordercolor={1 .7 .7}}
\providecommand{\arxivref}[2]{\href{http://arxiv.org/abs/#1}{#2}}

\providecommand{\href}[2]{#2}
\providecommand{\arxivlink}[1]{\href{http://arxiv.org/abs/#1}{arxiv:#1}}

\newcommand{\PSI}{F}

\newcommand{\comment}[1]{{}}

\newcommand{\su}{\frak{su}}

\newcommand{\mx}{x^{\rm mir}}

\newcommand{\ph}{{\rm ph}}
\newcommand{\mir}{{\rm mir}}
\newcommand{\rb}{\right)}
\newcommand{\lb}{\left(}
\newcommand{\nn}{\nonumber}

\newcommand{\Q}{{\cal Q}}
\newcommand{\X}{{\cal X}}
\newcommand{\Y}{{\cal Y}}
\newcommand{\beq}{\begin{equation}}
\newcommand{\eeq}{\end{equation}}
\newcommand\beqa{\begin{eqnarray}}
\newcommand\eeqa{\end{eqnarray}}
\newcommand\bea{\begin{array}}
\newcommand\eea{\end{array}}

\newcommand\IM{{\rm Im}\,}

\def\XXint#1#2#3{{\setbox0=\hbox{$#1{#2#3}{\int}$}
\vcenter{\hbox{$#2#3$}}\kern-.5\wd0}}

\newcommand{\COMMENT}[1]{}

\newcommand{\neqa}{\nonumber\end{eqnarray}}
\newcommand{\la}[1]{\label{#1}}

\newcommand{\eq}[1]{(\ref{#1})}

\newcommand{\matr}[2]{\left(\begin{array}{#1}#2\end{array}\right)}
\newcommand{\smatr}[4]{\matr{c|c}{#1&#2\\\hline #3&#4}}

\newcommand{\<}{{\langle}}
\renewcommand{\>}{{\rangle}}

\newcommand{\re}{\relax{\rm I\kern-.18em R}}

\def\su2{{SU(2)}}

\def\[{\left[}
\def\]{\right]}

\def\e{\epsilon}

\def\s{\sigma}

\def\<{\langle}
\def\>{\rangle}

\def\i2{\frac{i}{2}}

\begin{document}

%%%%%%%%%%%%%%%%%%%%%%%%%%%%%%%%%%%%%%%%%%%%%%%%%%%%%%%%%%%%%%%%%%%%%%%%%%%%%%%%
%%%%%%%%%%%%%%%%%%%%%%%%%%%%%%%%%%%%%%%%%%%%%%%%%%%%%%%%%%%%%%%%%%%%%%%%%%%%%%%%
% TITLE PAGE

\thispagestyle{empty}
%\phantomsection
\addcontentsline{toc}{section}{Title}

\begin{flushright}\footnotesize%
\texttt{kcl-mth-10-22},
\texttt{\arxivlink{1012.3996}}\\
overview article: \texttt{\arxivlink{1012.3982}}%
\vspace{1em}%
\end{flushright}

\begingroup\parindent0pt
\begingroup\bfseries\ifarxiv\Large\else\LARGE\fi
\hypersetup{pdftitle={Review of AdS/CFT Integrability, Chapter III.7: Hirota Dynamics for Quantum Integrability}}%
Review of AdS/CFT Integrability, Chapter III.7:\\
Hirota Dynamics for Quantum Integrability
\par\endgroup
\vspace{1.5em}
\begingroup\ifarxiv\scshape\else\large\fi%
\hypersetup{pdfauthor={Nikolay Gromov, Vladimir Kazakov}}%
Nikolay Gromov$^a$,  Vladimir Kazakov$^b$
\par\endgroup
\vspace{1em}
\begingroup\itshape
$^a$ King's College, London Department of Mathematics WC2R 2LS, UK\\
St.Petersburg INP, St.Petersburg, Russia\\
$^b$ Ecole Normale Superieure, LPT,  75231 Paris CEDEX-5, France\\
l'Universit\'e Paris-VI, Paris, France
\par\endgroup
\vspace{1em}
\begingroup\ttfamily
$^a$nikolay.gromov@kcl.ac.uk
\par\endgroup
\vspace{1.0em}
\endgroup

\begin{center}
\includegraphics[width=5cm]{TitleIII7.mps}%figure for your chapter
\vspace{1.0em}
\end{center}

\paragraph{Abstract:}
We review recent applications of the integrable discrete Hirota dynamics (Y-system)
in the context of calculation of the  planar AdS/CFT spectrum.
We start from the description of solution of Hirota equations by the B\"acklund method
where the requirement of analyticity results in the nested Bethe ansatz equations.
Then we discuss  applications of the Hirota dynamics for the analysis of the asymptotic limit of long operators in the AdS/CFT Y-system.

\ifarxiv\else
%\paragraph{Mathematics Subject Classification (2010):}
%...
% http://www.ams.org/msc
\fi

\ifarxiv\else
%\paragraph{Keywords:}
%...
\fi

\newpage

%%%%%%%%%%%%%%%%%%%%%%%%%%%%%%%%%%%%%%%%%%%%%%%%%%%%%%%%%%%%%%%%%%%%%%%%%%%%%%%%
%%%%%%%%%%%%%%%%%%%%%%%%%%%%%%%%%%%%%%%%%%%%%%%%%%%%%%%%%%%%%%%%%%%%%%%%%%%%%%%%
% BODY

%%%%%%%%%%%%%%%%%%%%%%%%%%%%%%%%%%%%%%%%%%%%%%%%%%%%%%%%%%%%%%%%%%%%%%%%%%%%%%%%
\section{Introduction}\label{sec:intro}

The Hirota integrable hierarchy   \cite{Hirota}  enables us to take  a very
general point of view on integrability,  in classical systems \cite{Jimbo:1983if}   as well as   in 2D statistical mechanical  \cite{Pearce:1991} and     quantum
 systems \cite{Zamolodchikov:TBA1990,Bazhanov:1994ft,Krichever:1996qd}.      The analytic Bethe ansatz  approach based on the Y- and T-systems for the fusion of transfer-matrices in various representations was successfully applied to various spin chains and 2D QFT's
 \cite{Krichever:1996qd,Bazhanov:1996aq}
and is especially efficient for the  supersymmetric systems
\cite{Tsuboi:1997iq,Kazakov:2007fy,Kazakov:2007na}. Being integrable, Hirota equation with specific boundary
conditions stemming from the symmetry of the problem,  can be often solved
explicitly,  either by the B\"acklund method \cite{Krichever:1996qd,Kazakov:2007fy}  or in the determinant (Wronskian) form \cite{Krichever:1996qd,Bazhanov:2008yc}.
Of course to specify completely the physical solutions we have to precise the functional space for the functions of spectral parameter entering Hirota equation, or in other words, we also need to impose certain analyticity
conditions on these solutions which is usually the hardest part of the problem. In the spin chains the role of analyticity conditions is usually played by the polynomiality of the transfer-matrices, resulting in supplementary
conditions
-  the Bethe ansatz equations. For the Y-systems of integrable 2D QFT's  (sigma models) at a finite volume, the analyticity imposes the absence of singularities on a physical domain of the complex plane of a spectral parameter, except those related to various physical excitations (see \cite{Gromov:2008gj} for the example of \(O(4)\) sigma model).

These methods, based on the Hirota integrable dynamics, recently have shown again their power in the problem of calculation of the exact conformal dimensions in the planar N=4 SYM theory. The program of integrability for the spectrum
in planar AdS/CFT correspondence has lead to the discovery of a system of exact spectral equations --- the Y-system ---
containing an important  information
about the anomalous dimensions of  all local operators at arbitrary 't Hooft coupling. The AdS/CFT Y-system and the underlying integrable Hirota equation were first conjectured   in a functional form   \cite{Gromov:2009tv} and later reproduced in the form of an infinite system of non-linear integral equations \cite{Bombardelli:2009ns,Gromov:2009bc,Arutyunov:2009ur} from the  TBA approach \cite{Zamolodchikov:TBA1990}. It was successfully tested analytically in the weak coupling regime, in particular  for  Konishi operator and
twist-2 operators by the direct
4-loop perturbation theory, and even up to 5 loops,  comparing with the BFKL approximation
 \cite{Bajnok:2008bm}, and in the strong coupling for long operators, by  comparison with
 quasi-classical string theory results \cite{Gromov:2009tq,Gromov:2010vb}. The first
 numerical study of Konishi dimension  \cite{Gromov:2009zb} in a wide range of couplings (see Fig.2 of
 \cite{RewIII6}), showed a
 perfect interpolation between the \(N=4\) SYM  perturbation theory and
 the SYM strong coupling asymptotics  described by the
 large radius of the superstring \(AdS_5\times S^5\) background   \cite{Gubser:1998bc}.

In this paper, we will introduce the reader into the basics of  Hirota
approach to the quantum integrability on the example of AdS\(_5\)/CFT\(_4\) duality. But first we  will show how to solve,  following the methods of
\cite{Kazakov:2007fy,Zabrodin:2007rq,Krichever:1996qd,Tsuboi:1997iq},
Hirota equation for the fusion in the
rational supersymmetric spin chains with  \(gl(N|M)\) symmetry,
in terms of a generating functional (generalized Baxter equations) by means of the B\"acklund method, and to derive the nested  Bethe ansatz equations. Then we will follow this logic in the AdS/CFT system
 and try to show that the worldsheet scattering theory and asymptotic Bethe ansatz  ABA for the superstring on  AdS$_5$/CFT$_4$ background are also tightly related to the analyticity of the Y-system  whose form, in its turn,  is greatly constrained by the superconformal \(psu(2,2|4)\) symmetry of the theory.

\section{Hirota Equations in the \texorpdfstring{\(gl(N)\)}{gl(N)}  spin chain}
The simplest example where the Hirota equation appears naturally is
the generalized Heisenberg \(gl(N)\)  spin chain for compact representations.
The spin chain Hamiltonian and all  other conserved charges can be
constructed from the R-matrix  \cite{Faddeev:1996iy}.
We give in this section the basics of the analytic Bethe ansatz approach to this system  following  \cite{Kazakov:2007fy,Kazakov:2007na,Zabrodin:2007rq}.

Our the explanations (though not  the proofs) will be self-contained,  all the way from the R-matrix to the Hirota equation.
The R-matrix of  \(gl(N)\)  super-spin chain is \begin{equation}
    R_{\lambda}(u)=u\,\,\mathbb{I}\  \otimes\mathbb{I}_{\lambda}+i\sum_{\alpha,\beta=1}^{N}e_{\beta\alpha}
     \otimes  \pi_{\lambda}(e_{\alpha\beta}) \,, \label{RMATRsup}
\end{equation}
\begin{figure}[ht]
\begin{center}
\includegraphics[scale=0.6]{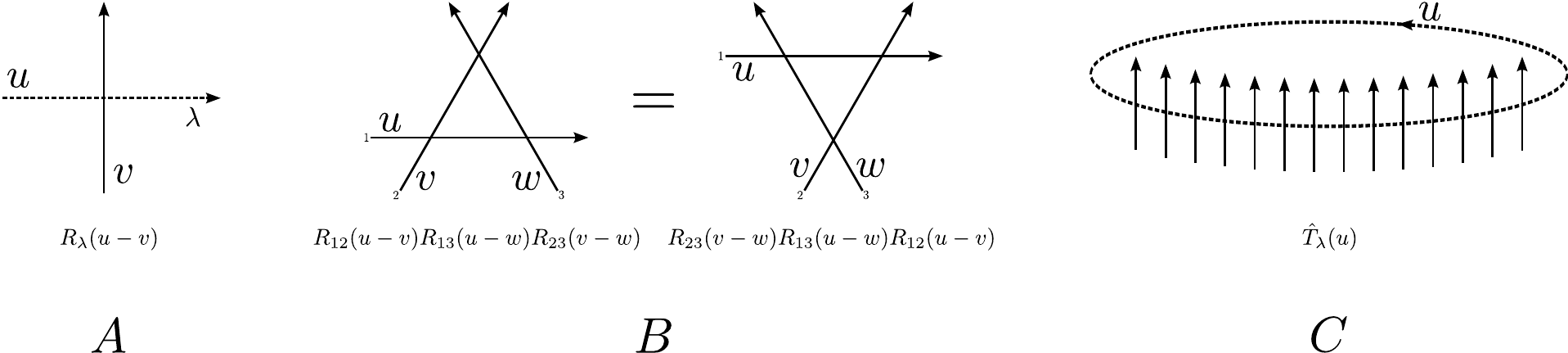}
\end{center}
\caption{A:  $R$-matrix in  pictures; B: Yang-Baxter relation; C: Transfer matrix
\label{fig:pics}}
\end{figure}
where the generators in the  l.h.s.(r.h.s.) of  the tensor product in each term correspond to the ``physical" (``auxiliary") space and
    $\lambda$ refers to an arbitrary  representation   in the ``auxiliary"
space.  \(\mathbb{I}\) and \(\mathbb{I}_{\lambda}\) are the identity elements
in fundamental representation and in  representation \(\lambda\), respectively; \(e_{\alpha\beta}\) are the generators of \(u(N)\) algebra acting in the fundamental representation
on the basis $e_\gamma$ as
\(
e_{\alpha\beta}e_\gamma = e_\alpha\delta_{\beta\gamma}\;,
\)
and \(  \pi_\lambda(e_{\alpha\beta})\) are the same generators in  any irrep \(\lambda\).
In
 \( \lambda\)  fundamental, the second term
\(
\mathcal{P}\equiv\sum_{\alpha\beta}e_{\beta\alpha}  \otimes  e_{\alpha\beta} \,,
\)
becomes simply the permutation operator
and it is easy to check that the R-matrix satisfies
the Yang-Baxter equation
\beq
R_{12}(u)R_{13}(u+v)R_{23}(v)=
R_{23}(v)R_{13}(u+v)R_{12}(u)
\eeq
where the operators act on the tensor product of $3$ fundamental physical  states
and the lower indexes show on which of the states the action of $R$ is nontrivial
(see Fig.\ref{fig:pics}B).

Next, we introduce the transfer matrix as a trace in the auxiliary space of irrep \(\l\) of the monodromy matrix (see the
Fig.\ref{fig:pics}C):
 \begin{equation*}
 \hat T_{\lambda}(u,g)\equiv{\rm tr_{aux}}\,
 \left(R_{\lambda}(u)^{\otimes L}\pi_{\lambda}(g)\right)\,,
\end{equation*}
where the tensor products are taken for the physical
spaces and the usual matrix product and the trace refers to the the auxiliary
space,
\(\pi_\lambda(g)\) being a group element $g$ in the irrep \(\l\).
The transfer matrix $\hat T_{\lambda}(u,g)$ is thus an operator
acting on $L$ copies of the physical space, i.e. on the Hilbert space of
the spin chain with $L$ cites. Notice that for $L=0$ the transfer matrix
is simply a character $\chi_{_\lambda}(g)$.

To relate the transfer matrices to the group characters, we introduce a useful operator called the  co-derivative \({\cal D}\)
\cite{Kazakov:2007na}
defined by the action on a function of  \(g\):
\begin{equation}
{\cal D}  f(g)=e_{\beta\alpha}  \left.\frac{\partial}{\partial \phi_{\alpha\beta}}
f\left(e^{ \phi_{\delta\gamma} e_{\delta\gamma}} g\right)\right|_{\phi=0}\,,\quad {\rm where}\qquad
\frac{\partial}{\partial \phi_{\alpha_1\beta_1}}\phi_{\alpha_2\beta_2} \equiv
\delta_{\alpha_1\alpha_2} \delta_{\beta_1\beta_2} \la{dsuper} \,.
\end{equation}
In particular, applying  it to  \eq{RMATRsup},  we rewrite the transfer matrix  in an instructive way
\begin{equation}\label{Lsup}
  \hat T_{\lambda}(u) = (u+i {\cal D})^{\otimes L}\chi_{_{\lambda}}(g)\;.
\end{equation}
In what follows we consider only the representations \(\l=s^a\) with rectangular Young diagrams $\lambda_i=s,\;i=1,\dots,a$.
Below we   demonstrate that the transfer matrices with
different spectral parameters $u$ and irreps $\lambda$ commute with each other
and thus we can work with their eigenvalues  denoted below as \(T_\lambda(u)\).
We denote \( \chi_{_{a,s}}\equiv\chi_{s^a}\)  and
\beq\label{eq:TdefS}
\hat T_{a,s}(u)\equiv\frac{\hat T_{_{s^a}}(u+\tfrac{s-a}{2i})}{(u+\tfrac{s-a}{2i})^L}
\eeq
where we chose the normalization of the eigenvalues
\begin{equation}\label{eq:norma}
T_{a,0}=T_{0,s}=1\;.
\end{equation}
The goal of this section is to demonstrate
the following Hirota equation \cite{Pearce:1991,Krichever:1996qd}:
\begin{equation}
\label{eq:HirotaKM}T_{a,s}(u+\tfrac{i}{2})T_{a,s}(u-\tfrac{i}{2})=T_{a+1,s}(u)T_{a-1,s}(u)+T_{a,s+1}(u)T_{a,s-1}(u)
\end{equation}

Let us demonstrate the validity of \eqref{eq:HirotaKM} on the case $a=1$.
The symmetric characters \(\chi_{_{1,s}}(g)\)
 are generated as the Schur polynomials  from the generating function
\begin{equation}\label{eq:genchar}\\ w(z)=\det(1-zg)^{-1}=\sum_{s=0}^\infty z^s \chi_{_{1,s}}.
\end{equation}
Acting on $w(z)$ by the left co-derivative we easily find that
\beq
{\cal D} \log w(z)=\frac{zg}{1-zg}
\eeq
\beq
(1+{\cal D})w(z_1){\cal D} w(z_2)=\frac{1}{1-z_1g}\frac{z_2g}{1-z_2g}=
\frac{z_2}{z_1}\frac{z_1 g}{1-z_1g}\frac{1}{1-z_2g}=
\frac{z_2}{z_1}\,{\cal D} w(z_1)(1+{\cal D})w(z_2)\;.
\eeq
The last equation in particular implies the following relation among the characters
\beq
{\cal D}\chi_{1,s}(\chi_s+{\cal D}\chi_s)={\cal D}\chi_{1,s+1}(\chi_{s-1}+{\cal D}\chi_{s-1})
\eeq
which, for the simple one spin chain \(L=1\), is equivalent to a particular case of \eqref{eq:HirotaKM}
\beq
\hat T_{1,s}(u+\tfrac{i}{2})
\hat T_{1,s}(u-\tfrac{i}{2})=
\hat T_{0,s}(u)
\hat T_{2,s}(u)+\hat T_{1,s-1}(u)\hat T_{1,s+1}(u)
\eeq
where we had to use that $\chi_{2,s}=\chi_{1,s}^2-\chi_{1,s+1}\chi_{1,s-1}$ and $\chi_{0,s}=1$.
Moreover, one can see that the one spin transfer matrices are a combinations of only
 $g$ and the unit matrix and thus commute with each other.
We send the interested reader to \cite{Kazakov:2007na} for the general proof of
the Hirota relation \eqref{eq:HirotaKM} for any irrep and any number of spins.
Eq.\eqref{eq:HirotaKM} is a generalization of a similar, but simplified, Hirota relation among the characters: \( \chi_{_{a,s}}^2=\chi_{_{a+1,s}}\chi_{_{a-1,s}}+\chi_{_{a,s+1}}\chi_{_{a,s-1}}\) -  following from the multiplication of rectangular irreps.

It is remarkable  that the fusion equation \eqref{eq:HirotaKM} is the same
for all $gl(N)$ groups. Different $N$ will correspond however to different boundary conditions.
In particular, one has \(T_{N+1,s}=0\)  (as well as \(T_{a<0,s}=T_{a\ne0,s<0}=0\)),  \(T_{a<0,s}=T_{a\ne0,s<0}=0\) which is clear from the same conditions for the characters: \(\chi_{N+1,s}=0\). It turns out that for the super groups $gl(N|M)$
the Hirota equation is again the same whereas the nonzero $T_{a,s}$ belong to so called fat-hook  \cite{Tsuboi:1997iq}
(see Fig.\ref{fig:fh}a).

It is easy to check that the ``gauge" transformation\footnote{We will  often use the notations
\( f^\pm=f(\theta\pm\frac{i}{2})\;,\,\, f^{\pm\pm}= f(\theta\pm
  i)\), and in general \(f^{[\pm k]}=f(\theta\pm\frac{i}{2}k)\).}
\beq\la{eq:gauge}
T_{a,s}\to g_1^{[a+s]} g_2^{[a-s]} g_3^{[s-a]} g_4^{[-a-s]} T_{a,s}
\eeq
where $g_i$ are arbitrary functions, leaves the form of the Hirota equation unchanged.
One may choose certain normalization of the solutions by fixing these functions in one or
another way, as we do in \eq{eq:norma}. Notice that \eq{eq:norma} still leaves one gauge
degree of freedom unfixed\footnote{Another normalization, more
natural for the spin chains, is to require $T_{a,s}(u)$ to be polynomial.
This corresponds to \eq{eq:TdefS} without denominator. For the AdS/CFT applications, and for the sigma-models in general,
these requirements of polynomiality are too strong}.
We can also introduce  the quantities
 gauge invariant w.r.t.  \eq{eq:gauge}  \beq\la{Ydef}
Y_{a,s}=\frac{T_{a,s+1}T_{a,s-1}}{T_{a+1,s}T_{a-1,s}}\;
\eeq
As a consequence of Hirota equations \eq{eq:HirotaKM} they satisfy the discrete Y-system equations
\begin{equation}
\label{eq:Ysystem} Y_{a,s}^+ Y_{a,s}^-
 =\frac{(1+Y_{a,s+1})(1+Y_{a,s-1})}{(1+1/Y_{a+1,s})(1+1/Y_{a-1,s})} \;.
\end{equation}

 There exists a  concise solution of Cauchy problem for Hirota equation in the semi-\((a,s)\)-plane , in terms of \(T_{1,s}(u)\) fixed along the boundary (recall that \(T_{0,s}(u)=1\) in our
 gauge), the so called Bazhanov-Reshetikhin determinant formula  \cite{Bazhanov:1987zu,Kazakov:2007na}
 for the fusion in spin chains (valid here in a more general context) \footnote{A similar formula expresses  \(T_{a,s}\)  through the antisymmetric characters \(T_{a,1}\). There exist also a generalization to the irreps with arbitrary Young tableaux.}
\begin{equation}\label{eq:BR}
T_{a,s}=\det_{1\le j,k\le a}T_{1,s+k-j}\left(u+\tfrac{k+j-a-1}{2i}\right)\;.
\end{equation}

  \begin{figure}
\begin{center}
\includegraphics[scale=0.42]{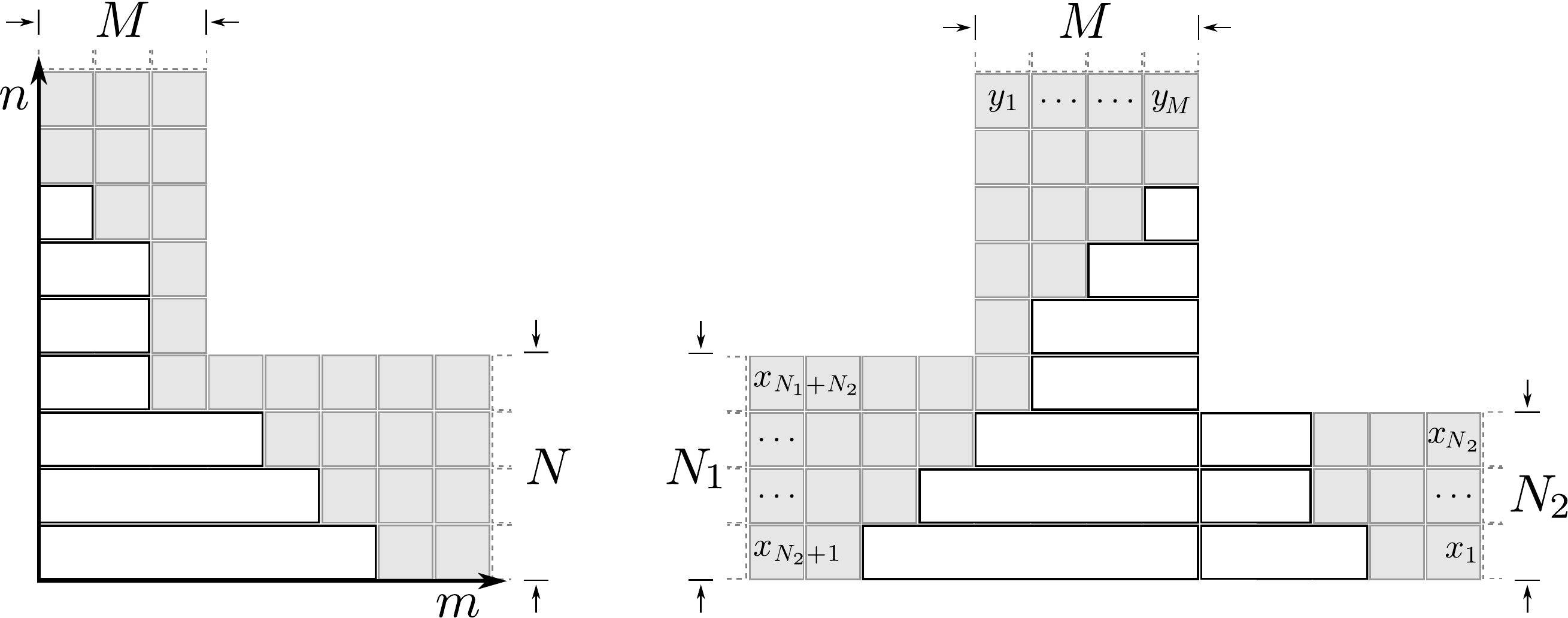}
\end{center}
\caption{The fat-hook for the representations of $SU(N|M)$ (left) and \(\mathbb{T}\)-hook for the representations of $SU(N_1,N_2|M)$ (right).
The lengths of  horizontal (white) strips forming the Young tableau of an irrep are equal to its highest weight components.\la{fig:fh}
}
\end{figure}

Our strategy will be to get as much information as possible about the system
by solving the Hirota equations.
The Bethe ansatz equations naturally appear in this approach
as a requirement of analyticity, or, in the case of spin chains, of  polynomiality of all transfer-matrices.
 In the next section we show how
 the Hirota  classical integrable discrete dynamics
 helps to solve, by means of the B\"acklund transform,  the fusion relations \eqref{eq:HirotaKM} in terms of a generating functional.

\section{Integrability of Hirota Equations}\la{sec:gen}

In this section we describe the general solution of Hirota equations
in the $(N|M)$ fat hook shown on the Fig.\ref{fig:fh}.
We apply for that the B\"{a}cklund transformation technique based on the classical integrability of discrete
 Hirota dynamics, and show how it helps to solve
the problem by gradually reducing the   \((N|M)\) fat hook to a trivial one \((0|0)\). As a result we derive the generating functional
for the general solution of Hirota equations.
In particular, the polynomial solution corresponds to the
 transfer matrices of the \(SL(N|M)\) rational Heisenberg super-spin chain
  described above.

\subsection{Linear system for Hirota equation}
The  classical integrability for the Hirota dynamics
manifests itself in the existence of an axillary linear problem - a pair of Lax equations
\begin{eqnarray}
\label{LINPRT1}
{\rm eq}^I_{a,s}(u)\;\;:\;\;  T_{a+1,s}\PSI^+_{a,s}
&=&+x\;T_{a+1,s-1}^+\PSI_{a,s+1}+T_{a,s}^+\PSI_{a+1,s}\;\;,\nn\\
{\rm eq}^{II}_{a,s}(u)\;\;:\;\; T_{a,s-1}\PSI^+_{a,s}
&=&-x\; T_{a+1,s-1}^+\PSI_{a-1,s}+T_{a,s}^{+}\PSI_{a,s-1}\;\;.
\end{eqnarray}
Their compatibility condition gives the Hirota equation \eqref{eq:HirotaKM}.
Indeed, we notice that $\PSI_{a,s}^{++}$ can be expressed
through $\PSI_{a+1,s-1},\;\PSI_{a,s},\;\PSI_{a-1,s+1}$
in two different ways:
1) use ${\rm eq}^I_{a,s}(u+\tfrac{i}{2})$ and then ${\rm eq}^{II}_{a,s+1}(u)$ with ${\rm eq}^{II}_{a+1,s}(u)$
2) use ${\rm eq}^{II}_{a,s}(u+\tfrac{i}{2})$ and then ${\rm eq}^{I}_{a-1,s}(u)$ with ${\rm eq}^{I}_{a,s-1}(u)$.
If we subtract the two results only the term linear in $x$ survives which implies:
\beq
\frac{T_{a,s}^{+}T_{a+1,s-2}}{T_{a,s-1} T_{a+1,s-1}^-}+\frac{T_{a,s}^{+}T_{a+2,s-1}^{}}{T_{a+1,s}T_{a+1,s-1}^-}-
\frac{T_{a+1,s-1}^{+}T_{a-1,s}}{T_{a,s-1}T^-_{a,s}}
-\frac{T_{a+1,s-1}^{+}T_{a,s+1}^{}}{T_{a+1,s}T_{a,s}^-}=0
\eeq
or, to put it differently, the function defined by
\(
f_{a,s}=\frac{T_{a-1,s}T_{a+1,s}+T_{a,s-1}T_{a,s+1}}{T_{a,s}^{+}T_{a,s}^-}
\)
should be periodic under the shift $f_{a,s}(u)=f_{a+1,s-1}(u)$. Since for the
transfer matrices  $T_{0,s}=1$ this implies that $f_{0,s}=1$ and thus \(f_{a,s}\equiv 1\),  leading to Hirota eq. \eqref{eq:HirotaKM}.

Next, noticing that the Hirota equation is invariant under $(a,s,u)\to (-a,-s,-u)$
we can easily find another linear system (useful for the next section)
\begin{eqnarray}
\label{LINPRT2}
T_{a-1,s}\tilde \PSI^-_{a,s}
&=&+y\;T_{a-1,s+1}^-\tilde\PSI_{a,s-1}+T_{a,s}^-\tilde\PSI_{a-1,s}\;\;,\nn\\
T_{a,s+1}\tilde\PSI^-_{a,s}
&=&-y\; T_{a-1,s+1}^-\tilde\PSI_{a+1,s}+T_{a,s}^{-}\tilde\PSI_{a,s+1}\;.
\end{eqnarray}

\subsection{Solution of Hirota fusion equation by the B\"acklund method}
As it was announced above the B\"acklund method allows to reduce the Hirota equation in a fat-hook $(N|M)$
to the same equation in a smaller fat hook $(n|m)$ with $n\leq N,\;m\leq M$.
For that we notice that \eq{LINPRT1}, after the  appropriate shifts in the spectral parameter and in $a$ and $s$,
can be written in the form
\begin{eqnarray}\label{lin12}
\PSI_{a-1,s}T^-_{a,s}
&=&+x\;\PSI^-_{a-1,s+1}T_{a,s-1}+\PSI^-_{a,s}T_{a-1,s}\;\;,\nn\\
\PSI_{a,s+1}T^-_{a,s}
&=&-x\; \PSI^-_{a-1,s+1}T_{a+1,s}+\PSI^-_{a,s}T_{a,s+1}
\end{eqnarray}
which is precisely the second linear system \eq{LINPRT2} with $\PSI_{a,s}$ and $T_{a,s}$ interchanged. In particular,
this implies that $\PSI_{a,s}$ should also satisfy the same Hirota equation. It is  always possible to choose $\PSI_{a,s}$
so that it satisfies Hirota equation in a smaller fat-hook $(N-1|M)$ i.e. to have
\(
\PSI_{N,s}=0\;\;,\;\;s>M.
\)
One can immediately see  from  \eqref{lin12}   that this condition is compatible with the fat hook boundary condition for T-functions \(T_{N+1,s>M}=0\).
Below we will construct this solution explicitly.

In view of this symmetry between $\PSI$ and $T$
we can denote
$T^{N|M}_{a,s}=T_{a,s}$
and $T^{N-1|M}_{a,s}\propto\PSI_{a,s}$ with a particular normalization \eq{eq:gauge}:
we normalize them so that $T_{0,s}=1$ and $T_{a,0}=1$. From \eq{LINPRT1}, this normalization implies  for $\PSI$ the following relations
\(
\PSI_{0,s+1}=\PSI^-_{0,s}\;\;,\;\;\PSI_{a-1,0}=\PSI^-_{a,0}
\)
which means that we can express $\PSI_{0,s}$ or $\PSI_{a,0}$ in terms of $\PSI_{0,0}$ with  a shifted  argument
\(
\PSI_{0,s}=\PSI_{0,0}(u-i\tfrac{s}{2}),\;\;\PSI_{a,0}=\PSI_{0,0}(u+i\tfrac{a}{2})\;.
\)
Thus in our normalization we get
\(
T_{a,s}^{N-1|M}\equiv\frac{\PSI_{a,s}(u)}{\PSI_{0,0}(u+\tfrac{s-a}{2i})}\;
\).
 It should be also clear that due to the symmetry
between $\PSI$ and $T$ we can change the logic and tell that \eq{LINPRT1} allows to increase
$M$.
Similarly, the second linear system \eq{LINPRT2} allows to decrease $M$ (or increase $N$) and
 we denote \(
T_{a,s}^{N|M-1}\equiv\frac{\tilde\PSI_{a,s}(u)}{\tilde\PSI_{0,0}(u+\tfrac{s-a}{2i})}\,
\).

By making an appropriate chain of these two transformations we can always reduce a fat hook ($N|M)$ to the trivial one $(0|0)$, through a set of the intermediate  fat hooks \((n|m), \quad 0<n<N; 0<m<M \) (see fig.\ref{fig:Path}).
This procedure allows to write the solution quite explicitly.
\footnote{This procedure is a ``quantum" analogue of the  construction of the so called Gelfand-Zeitlin basis.}
For the next section we introduce the parameterization
\beq\label{eq:psi_ident}
\frac{\PSI_{0,0}^{--}}{\PSI_{0,0}}\,=\frac{{\cal Q}_{N|M}^{++}{\cal Q}_{N-1|M}^{--}}{{\cal Q}_{N|M}{\cal Q}_{N-1|M}}
\;\;,\;\;
\frac{\tilde\PSI_{0,0}^{++}}{\tilde\PSI_{0,0}}\,=\frac{{\cal Q}_{N|M}^{--}{\cal Q}_{N|M-1}^{++}}{{\cal Q}_{N|M}{\cal Q}_{N|M-1}}
\eeq
The above equations define $\Q_{N|M-1}$ and $\Q_{N-1|M}$ in
terms of $\Q_{N|M}$, for  given functions \(\PSI_{0,0},\tilde\PSI_{0,0}\).
Since our normalization \eq{eq:norma} allows for one more gauge
one can set $\Q_{N|M}$ to $1$. This, however, is not
the most convenient choice. In the case of  spin chains (as in section 2)
a natural chose is $\Q_{N|M}=\prod_{j=1}^L(u-\theta_j)$. In this normalization, at an arbitrary step, or nesting level \((n|m)\) of our B\"acklund procedure,
$\Q_{n|m}$ will be a polynomial,  the denominator of the rational functions $T_{a,s}^{n|m}(u-\tfrac{s-a}{2i})$ (like in \eq{eq:TdefS}).

Furthermore we denote
\beqa\la{eq:XQ}
\X_{n|m}=x_{n}\frac{{\cal Q}_{n|m}^{++}{\cal Q}_{n-1|m}^{--}}{{\cal Q}_{n|m}{\cal Q}_{n-1|m}}
\;\;,\;\;\Y_{n|m}=y_{m}\frac{{\cal Q}_{n|m}^{--}{\cal Q}_{n|m-1}^{++}}{{\cal Q}_{n|m}{\cal Q}_{n|m-1}}\;.
\label{eq:paramQ}\eeqa

\subsection{A recurrent equation for the generating functional}\label{ssec:GeNFun}
For the quantum generalization of the generating function for the characters \eqref{eq:genchar} we introduce
an operator valued functional\beq\label{eq:genFdef}
{\cal W}^{n|m}=\sum_{s=-\infty}^\infty D^s T^{n|m}_{1,s}(u)D^s
\eeq
where $D$
is a shift operator defined by \(Df(u)=f(u-\tfrac{i}{2})D\).
From \eq{LINPRT1} at $a=0$ we get in  notations \eq{eq:XQ}
\begin{eqnarray}
T^{n-1|m}_{1,s}(u)&=&T^{n|m}_{1,s}(u)-
{\cal X}_{n|m}(u+\tfrac{s-1}{2i})\;T_{1,s-1}^{n|m}(u+\tfrac{i}{2})
\end{eqnarray}
 where we introduced  new \(n|m\) indices for \(\PSI\) characterizing the ``level" on which we make this B\"acklund transformation. This implies
\beq\label{eq:n-shift}
{\cal W}^{n-1|m}={\cal W}^{n|m}\lb 1- D{\cal X}_{n|m}D\rb
\eeq
and similarly
\beq\label{eq:m-shift}
{\cal W}^{n|m}={\cal W}^{n|m-1}\lb 1-D
{\cal Y}_{n|m}D\rb\;.
\eeq

\begin{figure}[ht]
\begin{center}
\includegraphics[scale=.6]{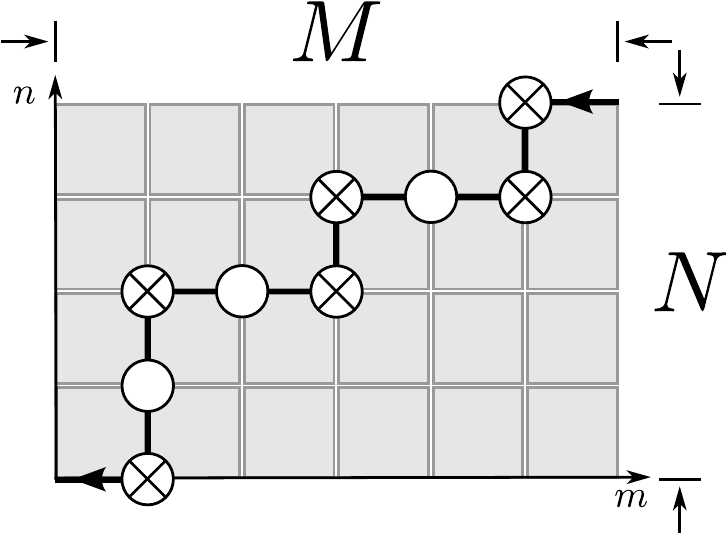}
\end{center}
\caption{B\"acklund procedure: a possible reduction path on the \((n|m)\)-lattice. }
\label{fig:Path}
\end{figure}

Using the relations \eqref{eq:n-shift} and \eqref{eq:m-shift} we can show that any solution of Hirota equation in the \((N|M)\) fat hook
can be explicitly and concisely written in the form of a simple generating functional. For that we have to apply the recursions  \eqref{eq:n-shift} and \eqref{eq:m-shift}
along a path of the length \(N+M\)   on the \((n|m)\)  lattice,  connecting  the upper right and the lower left corners of the  \(N\times M\)  rectangle on
Fig.\ref{fig:Path}. This gives
the following formula for the generating functional \eqref{eq:genFdef}
\cite{Krichever:1996qd,Tsuboi:1997iq,Kazakov:2007fy}
\beq\label{eq:GenFunl}
{\cal W}^{N|M}=\overleftarrow{\prod_{{\rm path}}}
\left\{
\bea{ll}
(1-D \X_{n|m} D)^{-1}&,\;\; {\rm vertical}\\
(1-D \Y_{n|m} D)&,\;\; {\rm horizontal}\\
\eea
\right.
\eeq
where the subset of $N+ M$ functions $\X_{n|m},\;\Y_{n|m}$, chosen out of the whole set of  $N\times M$ such functions,
depends on the path (see Fig.\ref{fig:Path}).
\beqa
\X_{n|m}\;\;,\;\;\text{line from}\;(n,m)
\;\text{to}\;(n-1,m)\\
\Y_{n|m}\;\;,\;\;\text{line from}\;(n,m)
\;\text{to}\;(n,m-1)
\eeqa
In the case when $\Q_{n|m}$ are polynomials the solution of  Hirota equation  constructed in this
way corresponds to the transfer matrices
with a twist given by a supergroup element \(g={\rm diag}\{x_1,\dots,x_N|y_1,\dots,y_M\}\).

\subsection{Analyticity and Bethe ansatz equations}
The parameterization of a solution of Hirota equation in terms of \(\Q\)-functions described above
is very convenient for constructing solutions with particular
analytic properties.  In this section we assume that ${\cal Q}_{n|m}$ are some functions without poles, more general than polynomials, which could have some zeros.
Generically,  in the process of the B\"acklund construction of a solution these zeros will create poles in the T-functions.
However it is possible to cancel all the poles in all \(T\)-functions by adjusting accordingly the
zeros of the ${\cal Q}_{n|m}$ functions. This will lead to a set of nested  Bethe ansatz equations.

To get it we notice that  if ${\cal Q}_{n|m}$ has a zero \(u^{(n|m)}_j\) then two vertical links in the path in fig.\ref{fig:Path}    meeting at the point  \((n|m)\) give the following (explicitly written) factor in the generating functional
\beqa
&&{\cal W}\simeq \dots \times(1-D\X_{n|m}D)^{-1}(1-D\X_{n+1|m}D)^{-1}\times\dots=
\\ \nn
&&\dots\times
\left[
1-D\lb
x_{n}\frac{{\cal Q}_{n|m}^{++}{\cal Q}_{n-1|m}^{--}}{{\cal Q}_{n|m}{\cal Q}_{n-1|m}}+
x_{n+1}\frac{{\cal Q}_{n+1|m}^{++}{\cal Q}_{n|m}^{--}}{{\cal Q}_{n+1|m}{\cal Q}_{n|m}}\rb D
+ x_k\,x_{n+1}\frac{{\cal Q}_{n+1|m}^{+}}{{\cal Q}^-_{n+1|m}}
D^4
\frac{{\cal Q}_{n-1|m}^{-}}{{\cal Q}^+_{n-1|m}}
\right]^{-1}\times\dots
\eeqa
In order to have no poles in \(T_{1,s}(u)\) we have to require that the poles \(u_j^{(n|m)}\) at  zeros of $\Q_{n,m}$ cancel in the round brackets giving the  Bethe ansatz equations
\beq\label{eq:nBA}
\left.\frac{{\cal Q}_{n+1|m}^{++}{\cal Q}_{n|m}^{--}{\cal Q}_{n-1|m}}
{{\cal Q}_{n+1|m}{\cal Q}_{n|m}^{++}{\cal Q}_{n-1|m}^{--}}\right|_{u^{(n|m)}_j}\!\!\!\!=-\frac{x_{n}}{x_{n+1}}\;\;,\;\;
\left.\frac{{\cal Q}^{--}_{n|m+1}{\cal Q}_{n|m}^{++}{\cal Q}_{n|m-1}}
{{\cal Q}_{n|m+1}{\cal Q}_{n|m}^{--}{\cal Q}^{++}_{n|m-1}}\right|_{u^{(n|m)}_j}\!\!\!\!=-\frac{y_{m}}{y_{m+1}}
\eeq
where the second equation comes from a similar cancelation for two neighboring horizonal links.
Similar cancelation can be seen for a horizontal link followed by a vertical  link meeting at  a point \((n|m)\)
\beqa
\nn &&(1-y_mD\Y_{n|m}D)(1-x_{n+1}D\X_{n+1|m}D)^{-1}\\
&=&\frac{{\cal Q}^{+}_{n|m-1}}{{\cal Q}^{-}_{n|m}}
(1-y_{m}D^2)
\frac{{\cal Q}^{+}_{n+1|m}}{{\cal Q}^{+}_{n|m-1}}
(1-x_{n+1}D^2)^{-1}
\frac{{\cal Q}^{-}_{n|m}}{{\cal Q}^{+}_{n+1|m}}\\&=&
\nn\lb\left[\frac{\Q^+_{n+1|m}}{\Q_{n|m}^-}-\frac{y_m}{x_{n+1}}\frac{{\cal Q}^{+}_{n|m-1}}{{\cal Q}^{-}_{n|m-1}}
\frac{\Q_{n+1|m}^-}{\Q_{n|m}^-}\right]
+\frac{y_m}{x_{n+1}}
\frac{{\cal Q}^{+}_{n|m-1}}{{\cal Q}^{-}_{n|m}}
\frac{\Q_{n+1|m}^-}{\Q_{n|m-1}^-}
\lb 1-x_{n+1}D^2\rb
\rb\\&\times&
(1-x_{n+1}D^2)^{-1}
\frac{{\cal Q}^{-}_{n|m}}{{\cal Q}^{+}_{n+1|m}}\;.
\nn
\eeqa
We see that we only have to require the cancelation of the poles in the square brackets
to ensure that the poles will not appear at any order in $D$. Similar relations can be written for a horizontal link following by a vertical one. That gives another pair of the Bethe equations, so that we have
\beq
\left.\frac{
\Q_{n|m-1}^{}{\cal Q}^{++}_{n+1|m}}{{\cal Q}^{++}_{n|m-1}\Q^{}_{n+1|m}}\right|_{u^{(n|m)}_j}\!\!=\frac{y_m}{x_{n+1}}\;\;,\;\;
\left.\frac{{\cal Q}^{}_{n-1|m}\Q_{n|m+1}^{--}}{\Q_{n-1|m}^{--}{\cal Q}^{}_{n|m+1}}\right|_{u^{(n|m)}_j}\!\!=\frac{x_n}{y_{m+1}}.
\eeq
Notice that this pair of equations is compatible with the first pair \eqref{eq:nBA}
-- their products coincide. This is a consequence of the ``zero curvature" equations discussed below. In general, we need only \(N+M-1\)
Bethe equations, written in the interior vertices of a path  of Fig.\ref{fig:Path}, to fix completely the full set of Q-functions with all their zeros.

\subsection{Self-consistency of the construction and \texorpdfstring{$QQ$}{QQ}-relations}\la{sec:du}
Once a path on Fig.\ref{fig:Path} is fixed one can choose an arbitrary set of
functions \({\Q_{n|m}}\) along this path in order to get some solution of the Hirota equation.
If  we want now  to change the nesting path  without changing the solution for T-functions, it is possible to choose a new subset of \(N+M-1\) \(\Q\)-functions
entering  the  generating functional \eqref{eq:GenFunl}.
Let us consider such an elementary modification of the functional:
\beq
(1-D\X_{n|m-1}D)^{-1}(1-D\Y_{n|m}D)=(1-D\Y_{n-1|m}D)(1-D\X_{n|m}D)^{-1}
\label{eq:DualSU22}\eeq
The terms quartic in \(D\)  cancel automatically and the quadratic terms give\beq
\X_{n|m-1}-\Y_{n|m}=\X_{n|m}-\Y_{n-1|m}
\label{eq:XYdual}\eeq
which means that the combination
\beq\la{eqQQ}
f_{n|m}=\frac{x_n\Q_{n-1|m-1}^-\Q_{n|m}^+-y_m\Q_{n-1|m-1}^+\Q_{n|m}^-}{\Q_{n-1|m}^-\Q_{n|m-1}^+}
\eeq
is a periodic function with a period \(i\). In the case when \(\Q_{n|m}\) are polynomials
\(f_{n|m}\) should be a constant, which leads to the following \(QQ\) relation \cite{Tsuboi:1997iq,Kazakov:2007fy}
\beq
f_{n|m}\,\,\Q_{n-1|m}^-\Q_{n|m-1}^+=x_n\,\Q_{n-1|m-1}^-\Q_{n|m}^+-y_m\,\Q_{n-1|m-1}^+\Q_{n|m}^-\,.
\eeq
In the spin chain case,  when $\Q$'s are polynomials, one can fix their
normalization to have the same lading large $u$ coefficient.
In this case, evidently $f_{n,m}=x_n-y_m$.

Now we will show how all these rather abstract considerations can help us to attack an important physical problem - the study of the Y-system for the exact spectrum of an AdS/CFT system.

\section{ Classical transfer matrix of  \texorpdfstring{\(AdS_5\times S^5\)}{AdS5xS5}   superstring  }

In this section, we remind the results of the finite gap solution of the classical superstring on \(AdS_5\times S^5\) \cite{Bena:2003wd,Beisert:2005bm}
(see  \cite{RewII4} for the details).
We will construct in the classical limit a set of the eigenvalues of transfer matrices in various representations.
We demonstrate that, very similarly to transfer matrices of the spin chains,
the classical transfer matrices (traces of the monodromy matrix in various irreps) of the Metsaev-Tseytlin sigma model satisfy the Hirota equation.
The crucial difference with the previous example is the non-compact  symmetry group $PSU(2,2|4)$
which implies a different type of boundary conditions for the Hirota equation - the so called ${\mathbb T}$-Hook (Fig.\ref{fig:fh}b).

We examine the properties of  solutions of Hirota eqs. given by the classical transfer matrices
and then in the next section we discuss certain aspects of the generalization to the quantum case.

\subsection{Characters of  \texorpdfstring{$PSU(2,2|4)$}{PSU(2,2|4)}   and their Hirota dynamics  }

The monodromy matrix $\Omega(x)$ is a spectral parameter dependent  \(SU(2,2|4)\) group element. In the fundamental representation it is a \(4|4\times4|4\) supermatrix with \(4+4\) eigenvalues \((x_1,\dots,x_4|y_1,\dots,y_4) \) expressed through the quasi-momenta (of  \(S^5\) and \(AdS_5\) respectively) as follows:   \(x_j=e^{-i\tilde p_j(x)}\,, y_j=e^{-i\hat p_j(x)}\,, j=1,2,3,4\).
The dependence of \(\Omega(x)\) on the spectral parameter \(x\)
comes from the expression for the Lax pair \cite{Beisert:2005bm}. Supertrace of the monodromy matrix   \(\Omega(x)\) in any unitary highest weight irreducible representation (irrep)
\(\lambda\) will be denoted by
\(T_{\lambda}={\rm Str}_{\lambda}\Omega(x)\)\footnote{Since all the unitary representations
are indefinite dimensional the supertrace may not be convergent in some special cases. For sufficiently large
$L/\sqrt\lambda$  the convergence is guarantied.}.

Such highest weight irreps of $U(2,2|4)$ can be parameterized by
 generalized Young diagrams (see Fig.\ref{fig:fh}b).
The rectangular irreps  $\lambda_i=s+2,\;i=1,\dots,a$ which we denote as \([a, s\)]
are playing a crucial role since they obey a closed system of relations w.r.t. their tensor  product
\(
[a, s]\otimes[ a, s]=[a+1, s]\otimes[a-1, s]\oplus[ a, s+1]\otimes[ a, s-1]\;
\).
Tracing out this relation we find that  the characters of  $T_{a,s}$
of such irreps again satisfy the Hirota relation, as it was the case for the characters \(\chi_{a,s}\) of the sec.2
\begin{equation}
 \label{Tsystem}
 T_{a,s}T_{a,s} =T_{a+1,s}T_{a-1,s}+T_{a,s+1}T_{a,s-1} \,.
\end{equation}
As we shell see later,  this equation is a special limit of the full quantum Hirota equation \eqref{eq:HirotaKM} containing no shift in the spectral parameter, since it is invisible in this system in the strong 't Hooft coupling $\lambda\to\infty$ limit
where the spectral parameter is parameterized as $u=\frac{\sqrt\lambda}{4\pi}(x+1/x)$ and scales as $\sqrt\lambda$.

Let us compare the characters for  finite dimensional irreps of \(U(4|4)\) and the characters of non-compact infinite dimensional irreps of \(U(2,2|4)\).
  They satisfy the same  Hirota equation \eqref{Tsystem} but
  with different boundary conditions in the infinite \((a,s)\) lattice.
Both are defined by the same generating function
\begin{equation}\label{GEN44}
w(z)
={\rm SDet}\left(1-z\,\Omega(x)\right)^{-1}=\frac{(1-y_{1}z)(1-y_{2}z)(1-y_{3}z)(1-y_{4}z)}
{(1-x_{1}z)(1-x_{2}z)(1-x_{3}z)(1-x_{4}z)}
\end{equation}
where the characters of irreps \((1,s)\) are generated by the contour integrals
\begin{equation}
T_{1,s}^{(4|4)}=\frac{1}{2\pi i}\oint_{C}\frac{dz}{ z^{s+1}}w(z)\;,
\label{oint44}\end{equation}
and all the other representations can be generated from
there irreps by
the Jacobi-Trudi type formula (which is a direct consequence of \eqref{Tsystem}):
\begin{equation}\label{eq:Jacobi-Trudi}
T_{a,s}=\det_{1\le i,j\le a}\, T_{1,s+i-j}\;.
\end{equation}
The two types of characters differ by the definition of the integration contour \(C\). If the contour encircles the origin, living aside all the poles in the denominator of    \eqref{GEN44}, then the corresponding \(T_{a,s} \) (also called the super-Schur polynomials) constructed from \(T_{1,s}\)  by means of \eqref{eq:Jacobi-Trudi}
will be non-zero only inside the so called \(4|4\) fat hook on the \((a,s)\)  lattice (see Fig.\ref{fig:fh}). This corresponds to the compact unitary representations of \(U(4|4)\). But if  the contour encircles the origin {\it\ together} with the poles
\(t=x_3^{-1},x_4^{-1}\)
 the corresponding characters generated by  \eqref{eq:Jacobi-Trudi} are non-zero only within  the \(\mathbb{T}\)-hook
Fig.\ref{fig:fh}b.
 It is shown in \cite{CLZ03,Kwon06} that the irreps corresponding to these characters are indeed the unitary infinite dimensional irreps of \(U(2,2|4)\) (see also \cite{Gromov:2010vb} for some explanations and for the  explicit formulas for these characters).

These characters have a few discrete symmetries.  They have a specific symmetry
w.r.t. to the inversion of the eigenvalues:
\begin{equation}\la{Tinv}
T_{a,s}(x_{1},\dots,x_{4}|y_{1},\dots,y_{4}) =
T_{a,-s}\left(\left.\frac{1}{x_{4}},\dots,\frac{1}{x_{1}}\right|\frac{1}{y_{4}},
\dots,\frac{1}{y_{1}}\right)
\end{equation}
and instead of the full Weyl symmetry of the compact irreps,  they have only a residual permutational symmetry
\begin{equation}\label{xperm} x_1,x_2\leftrightarrow x_2,x_1\;\;;\;\;
 x_3,x_4\leftrightarrow x_4,x_3\;\;;\;\; \{y_1,y_2,y_3,y_4\}\leftrightarrow{\rm Perm}\{y_1,y_2,y_3,y_4\}\;.
\end{equation}

They also have some complex conjugation properties
 described below.

\subsection{\texorpdfstring{$Z_4$}{Z4} symmetry and reality}

From the unitarity of the classical monodromy matrix, the eigenvalues as functions of  $x$ are unimodular
\beq\label{eq:xy-reaplity}
\overline{ x_i(x)}=1/x_i(\bar x)\;\;,\;\;\overline{ y_i(x)}=1/y_i(\bar x)\;.
\eeq

 The \(Z_4\)-symmetry
of this \(AdS_5\times S^5\)    coset model imposes the following  monodromy property
\comment{
\begin{equation}
C^{-1}
\Omega(x)C =
\Omega^{ST} (1/x)\;\;,\;\; {\rm where}\;\;
C=\smatr{E}{0}{0}{-iE},
\quad E=\smatr{0}{ 1}{- 1}{0}_{4\times 4}
\end{equation}
which means that the eigenvalues get reshuffled under the inversion of the
spectral parameter}  \cite{Beisert:2005bm}
\begin{equation}\label{eq:Z4}
x_{1,2,3,4}(1/x) =\frac{1}{x_{2,1,4,3}(x)}\;\;,\;\; y_{1,2,3,4}(1/x)
=\frac{1}{y_{2,1,4,3}(x)}.
\end{equation}  Since on the unit circle \(|x|=1\)  we have $\bar x=1/x$  and
we get
\beq
\overline{ x_{1,2,3,4}(x)}=\frac{1}{x_{1,2,3,4}(1/x)}=x_{2,1,4,3}(x)\;\;,\;\;\overline{ y_{1,2,3,4}(x)}
=y_{2,1,4,3}(x).\la{Xreality}
\eeq
All this, together with \eq{xperm}, implies the reality of \(T_{a,s}\) on the unit circle  \(|x|=1\):
\beq\label{eq:realT}
\overline{T_{a,s}}={T_{a,s}}\,.\eeq
Then the $Y$ functions defined by \eq{Ydef}
are also real: \(\overline{Y_{a,s}}={Y_{a,s}}\). It follows from the definition \eqref{Ydef} and the explanations below the eq.\eqref{eq:Jacobi-Trudi} that whereas \(T_{a,s}\) are non-zero
in the vertices of the  Fig.\ref{T-Hook}(left) the
 \(Y_{a,s}\) are defined only in the visible nodes on the Fig.\ref{T-Hook}(right). As was explained in
\cite{Gromov:2009tq} eq.\eqref{Tsystem} and the corresponding simplified  Y-system describe the quasi-classical limit of the AdS\(_5\)/CFT\(_4\) system.

\section{ Quantum Hirota equation for AdS/CFT}

There is no rigorous
prove that the Metsaev-Tseytlin (MT) superstring  \(\s\)-model
is a well defined quantum theory, though the explicit perturbative
 SYM calculations lead to  the results consistent up to two loops with the classical limit of MT model \cite{Serban:2004jf,Beisert:2005di}.
We know  that this \(\s\)-model is  classically integrable and that there is also an
abundant evidence
of its quantum mechanical  integrability.
The experience from relativistic   quantum \(\s\)-models with massive spectra shows that the problem of the energy spectrum on a finite space circle, or a finite radius 2D space-time cylinder,  always boils down to a
very simply looking and universal  system of functional  Y- and T-systems, or Hirota equations \eq{eq:HirotaKM},
 the same as for the spin chains considered in the Sec.2.
The boundary conditions in \(a,s\) and the analyticity conditions in \(u\)
for the Hirota-type system or the corresponding Y-system differ from model to model, but
usually their general form
\eq{eq:HirotaKM} is  tightly related to the underlying symmetry and stays the same for all \(gl(N|M) \) algebras (with only  minor modifications for other algebras)\footnote{For  an incomplete table of  integrable models
and their Y-systems see the last page of \cite{Gromov:2008gj}.}.
Unless there exists an integrable lattice version,
the only tangible proof of the Y-system for
each particular finite size \(\s\)-model
is based on the TBA approach \cite{Japaridze:1984dz}  with the finite temperature interpreted as a finite space circle  \cite{Zamolodchikov:TBA1990}.

The  quantum MT \(\s\)-model in the light-cone
gauge, looking as a massive, though not explicitly relativistic theory,
seems to be in the same class of integrable \(\s\)-models as the above mentioned relativistic examples.
The absence of the worldsheet relativistic invariance, necessary to swap the worldsheet time and space directions, complicates but does not ruin the TBA approach to the finite size problem.

\begin{figure}[ht]
\begin{center}
\includegraphics[scale=0.7]{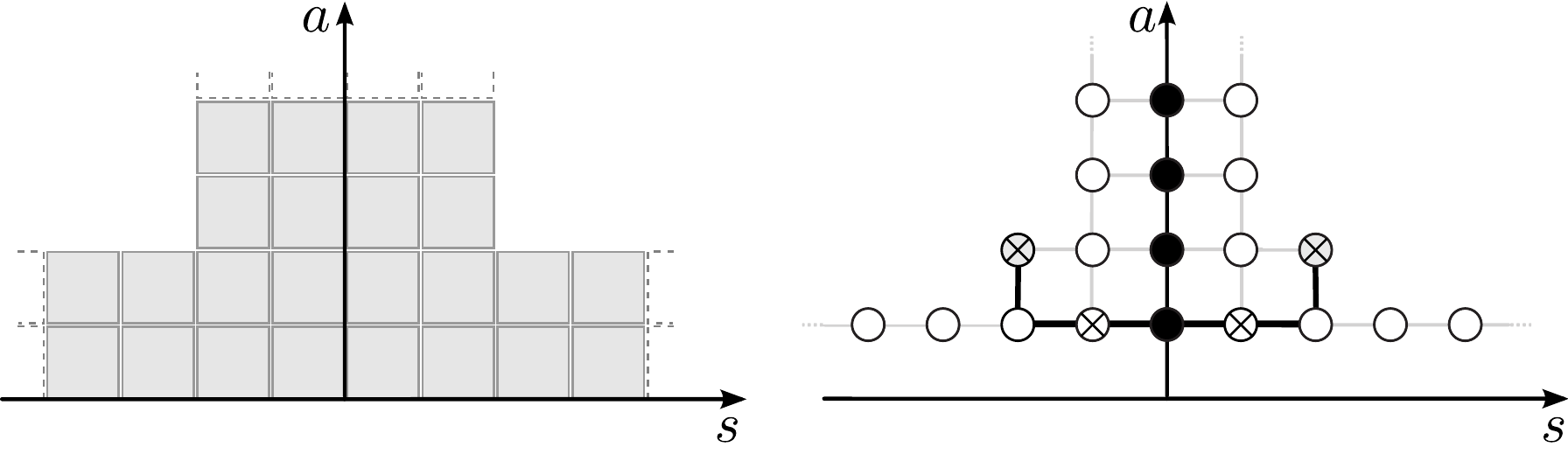}
\end{center}
\caption{\(\mathbb{T}\)-shaped ``fat hook" (\(\mathbb{T}\)-hook) uniting two ${\rm SU}(2|2)$ fat hooks,
see \cite{Gromov:2009tv} for this \(\mathbb{T}\)-hook and its generalization \cite{Hegedus:2009ky}.
}\label{T-Hook}
\end{figure}

To apply the T-system for a particular \(\s\)-model
one should identify the boundary conditions on the \((a,s)\)-lattice.
The quasi-classical picture of the previous section   suggests
that the full quantum Hirota equation should have the same boundary conditions, the  \(\mathbb{T}\)-hook of
Fig.\ref{T-Hook}, as the simplified
system for characters
\eqref{Tsystem}, as a consequence of the AdS/CFT superconformal \(PSU(2,2|4)\) symmetry.

The next step is to identify  the  spectral parameter \(u\) entering the full quantum Hirota eq.\eqref{eq:HirotaKM}. In analogy with the   integrable sigma-models   \cite{Dorey:2006mx}  it can be taken the same as entering the pair    \((p,u), \) where \(p\) is the quasi-momentum of the classical monodromy matrix,  defining  the simplectic structure of the algebraic curve and  entering the   holomorphic integrals \(\oint pdu\) of the Bohr-Sommerfeld quantization.      This parameter is related to the  one used  in the previous section    by   Zhukovski map  \cite{Beisert:2005bm}
\begin{equation}\label{eq:Zhuk}u=\tfrac{\sqrt\lambda}{4\pi}(x+1/x)\;.
\end{equation}

We will assume that this spectral parameter  \(u\)  is the same as in the full quantum AdS/CFT Y-system \eqref{eq:Ysystem}.
The initial spectral parameter \(x\) is then a double valued function w.r.t. the new parameter \(u\).
As a consequence of these   additional analyticity features in this construction we  expect that $Y_{a,s}$ has several cuts
parallel to the real axes, with the branchpoints   at $\pm 2g+\tfrac{in}{2}$.
To fix the cut structure we distinguish two kinematics: the physical and the mirror (where the  role of time and space is swapped)
\beq\la{xxs}
\qquad x^\ph(u)=\frac{1}{2}\lb \frac{u}{g}+\sqrt{\frac{u}{g}-2}\;\sqrt{\frac{u}{g}+2} \rb
\;\;,\quad\;\;\mx(u)=\frac{1}{2}\lb \frac{u}{g}+i\sqrt{4-\frac{u^2}{g^2}}\rb \,,
\eeq
having  branch cuts at  \((-2g,2g)\) and \((-\infty,-2g)\cup(2g,\infty)\), respectively.\footnote{\(x(u) \) as analytic function has the following conjugation properties: \(\bar x(u)=\frac{1}{x(u)}\) on the mirror sheet and \(\tilde x(u)=x(u)\) on the physical sheet.}

In the supersymmetric models the Hirota equation \cite{Gromov:2009tv}
appears to be a little more then the Y-system \eqref{eq:Ysystem} in which  two corner equations are missing.
When one tries truncate the Y-system from the full \((a,s)\) plane lattice to the \(\mathbb{T}\)-hook one has to put \(Y\)-functions to zero on the vertical boundaries, and to \(\infty\) at the horizontal boundaries in the left figure of
Fig.\ref{T-Hook}. Then the equation for
\(Y_{2,\pm 2}\)
contains an uncertainty \(\frac{0}{0}\). The Y-system has to be supplemented by  additional information.
In this respect, the T-system, free of that uncertainty, looks more fundamental than the Y-system.

 To fix the functions
\(Y_{2,\pm 2}(u)\) at
the corner nodes we will use a fact noticed from TBA \cite{Bombardelli:2009ns,Gromov:2009bc,Arutyunov:2009ur} (and partially inspired by ): in the mirror kinematics,
\(Y_{2,\pm 2}(u)\) and
\(Y_{1,\pm 1}(u)\) are related on two sides of the  cut \((-\infty,-2g)\cup(2g,\infty)\) on \(\mathbb{R}\) by

\begin{equation}\label{eq:Y11Y22}
Y_{2,\pm 2}(u+i0)=\frac{1}{Y_{1,\pm 1}(u-i0)}\,.
\end{equation}In the next section we will use  \eqref{eq:Y11Y22} as a natural analytic input for the asymptotic large \(L\) solution of the quantum Y-system.

Given a particular solution of the Y-system, the corresponding energy of a string state  (or anomalous dimension of  a SYM operator) can be obtained from\footnote{which can be partially motivated by a similar formula for the wrapping contributions in the quasiclassical quantization from the algebraic curve of the finite gap method, see \cite{Gromov:2008,Gromov:2010vb}.}
\begin{equation}
E=\sum_{j}\epsilon^{\ph}_1(u_{4,j})+\sum_{a=1}^\infty\int_{-\infty}^{\infty}\frac{du}{2\pi i}\,\,\frac{\partial\epsilon_a^{\rm mir}}{\partial u}\log\left(1+Y_{a,0}(u)\right) , \label{eq:Energy}
\end{equation}
whose general form is rather standard in the TBA context. The physical energy \(\e_a^{\rm ph}\) or the mirror momentum \(\epsilon_a^{\rm mir}\) are defined by the same formula \begin{equation}\label{eq:energyU}
\epsilon_a(u)= a+\frac{2ig}{x^{[+a]}}-\frac{2ig}{x^{[-a]}}\;.
\end{equation}
with the corresponding choice of $x(u)$ from \eq{xxs}.

The physical roots \(u_{4,j}\)   are subject to the exact, finite size  Bethe ansatz equations\beq\label{eq:Broots}
Y^\ph_{1,0}(u_{4,j})=-1\,
\eeq
where the $Y^\ph_{1,0}(u)$ was explicitly defined in \cite{Gromov:2009zb}
as an analytic continuation of $Y_{1,0}(u)$
down through the cut \((-\infty,-2g+\tfrac{i}{2})\cup(+2g+\tfrac{i}{2},+\infty)\). The first term  in \eqref{eq:Energy}  is given by the logarithmic pole contributions from the second one at the points \(u_{4,j}\).

It is also important to mention that at large \(L\) the  Y-functions of the middle (black) nodes are exponentially suppressed on the real axis in the mirror sheet
\begin{equation}\label{eq:wrapY}
Y_{a,0}(u)\sim e^{-ip^\mir_a(u)L}\,,\qquad {\rm where }\quad p_a(u)=-i\log\left(\frac{x^{[+a]}}{x^{[-a]}}\right)^L\;.
\end{equation}

\subsection{Integrability of  AdS/CFT Y-system and  large volume limit}

To study the AdS/CFT Y-system we need to clarify the analyticity properties of the Y-functions.
 Most of this information  is due to the
TBA derivation of the \(Y\)-system. The full understanding of these properties still needs additional efforts (see \cite{Cavaglia:2010nm} for some advances). We
will try to summarize them and demonstrate their naturalness. Ideal would be to postulate these properties from some simple and natural physical principles and then deduce from them the asymptotic Bethe ansatz (ABA) equations, along with the dressing factor (ignoring the standard S-matrix bootstrap procedure) as  it can be done for  various relativistic sigma-models    (see \cite{Gromov:2008gj,Kazakov:2010kf} for an inspiring example of the \(SU(N)\times SU(N)\) principal chiral field). On our current level of understanding of the AdS/CFT Y-system, this program can be fulfilled only partially.

This  Y-system is equivalent to   Hirota eq.\eqref{eq:HirotaKM} in the \(\mathbb{T}\)-hook fig.\ref{T-Hook}(left) with specific analyticity conditions.
Fortunately, many of the results for the simplified Hirota eq.\eqref{Tsystem} for quasi-classical AdS/CFT,  in particular \eqref{eq:Jacobi-Trudi} and \eqref{GEN44}, as well as the analyticity  \eqref{eq:xy-reaplity}-\eqref{eq:realT}, can be generalized to the full quantum case. We will demonstrate in this section that the asymptotic Bethe ansatz (ABA) of \cite{Beisert:2006ez} can be explained, and  partially  derived from the AdS/CFT Y-system \eqref{eq:Ysystem} together with the relation \eqref{eq:Y11Y22}, providing the reality of Y-functions, the \(s\leftrightarrow-s\) symmetry   and certain natural  analyticity assumptions, such as the existence of analyticity strips in \(u\)-plane.
\subsection{Generating functional for \texorpdfstring{\({\rm U}(2,2|4)\)}{U(2,2|4)} T-functions}
Since  Hirota equation for AdS\({_5}\)/CFT\({_4}\)  is exactly the same as the one considered in the Sec.2 for  spin chains
one may try to construct its general solution in terms of only a few functions. But we here we deal with a non-compact symmetry group, and the
 ${\mathbb T}$-hook instead of the usual ${\mathbb L}$-shaped fat hook domain for the Y-system as a consequence.
 In the pervious section, in the strong coupling limit  the difference between  the $U(4|4)$
 and $U(2,2|4)$  generating functions was only in  the way we expand  various parts of   \eq{GEN44} w.r.t. the generating parameter \(t\).
 A  natural generalization of  \eq{eq:GenFunl}  for the quantum case or the \(\mathbb{T}\)-hook  gives \(T_{1,s}(u)\) in terms of the generating functional \cite{Gromov:2010vb}
\beqa\label{eq:GenFun224}
W&=&\left[(1-D\Y_1 D)
\frac{1}{1-D\X_1 D}
\frac{1}{1-D\X_2 D}
(1-D\Y_2D)\right]_+ \times\\
\nn&&\left[(1-D\Y_{3}^{}D^{})
\frac{1}{1-D^{}\X_{3}^{}D^{}}\frac{1}{1-D^{}\X_{4}^{}D}
(1-D\Y_{4}D)\right]_-
=\sum_{s=-\infty}^{\infty} D^{s}T_{1,s}D^{s}
\eeqa
Here \(\{\Y_1(u)|\X_1(u),\X_2(u)|\Y_2(u),\Y_3(u)|\X_3(u),\X_4(u)|\Y_4(u)\}\) are 8 arbitrary functions of the spectral parameter \(u\) parameterizing the general solution
where, as a convenient choice for the AdS/CFT system, the grading is fixed by the Kac-Dynkin diagram
\(\bigotimes\!\!-\!\!\bigcirc\!\!-\!\!\bigotimes\!\!-\!\!\bigcirc\!\!-\!\!\bigotimes\!\!-\!\!\bigcirc\!\!-\!\!\bigotimes\).
Similarly to  the \(U(2,2|4) \) characters  (see after  eq. \eq{eq:Jacobi-Trudi}),
we expand in positive powers of the shift operator    \(D\)~~\footnote{We remind that $D$ is defined by $D f(u)=f(u-i/2)D$.
Since presently we may expect branch cuts originated from the map $x(u)$ the shift may be ambiguous, the prescription
is to analytically continue along the path going between the branch points without crossing the cuts going to infinity parallel to the real axes.}
 (replacing the \(t \) of \eqref{GEN44}) inside the bracket \(\left[\dots\right]_+\) corresponding to the \(u(2|2)_R\) sub-algebra, and in negative powers of \(D\) inside the bracket \(\left[\dots\right]_-\) corresponding to the \(u(2|2)_L\) subalgebra
  \footnote{As was noticed in \cite{Beisert:2006qh}, we can generate symmetric and antisymmetric representations by expanding the generating functional in powers of \(D\) and \(D^{-1}\) respectively. Mixed expansions generate infinite representations for non-compact real forms of \(gl(M|N)\). }.
As a result one gets an infinite sum for each $T_{1,s},\,\,\,-\infty<s<\infty$. Note also that \eqref{eq:GenFun224} corresponds
to a gauge where \(T_{0,s}=1\) and all  other $T_{a,s}$ can be found  from \eq{eq:BR}.

In the asymptotic $L\to\infty $ limit the full Y-system in the \(U(2,2|4)\) \(\mathbb{T}\)-hook almost splits into two Y-subsystems of two
$su(2|2)_{L,R}$ fat hooks  corresponding to the \(L,R\) wings:
$\X_1,\X_2,\Y_1,\Y_2$ are exponentially small whereas $\X_3,\X_4,\Y_3,\Y_4$
are exponentially large, thus the terms in the sum over \(s\) are organized
in  powers of  wrapping\footnote{Wrappings are related to the Feynman graphs wrapped around the ``spin chain" representing an operator of a length \(L\): in weak coupling, \(k\) wrappings occur at the order \(\lambda^{Lk}\)}.

We can easily find
these 8 functions in the $L\to\infty $ limit by comparing \(T_{1,1}(u)\) generated from \eqref{eq:GenFun224}
with the explicit asymptotic solution of the Y-system with given Bethe roots found  in \cite{Gromov:2009tv}
 (partially by matching with the known ABA of  \cite{Beisert:2006ez}) \begin{eqnarray}\label{eq:Q-monom}
\Y_{1}=H_R F_0^+ \frac{\Q_1^{-}}{\Q_1^+}\;\;,\;\;
\X_1=H_R\frac{\Q_1^{-}\Q_2^{++}}{\Q_1^+ \Q_2 } \;\;,\;\;
\X_2=H_R\frac{\Q_2^{--}\Q_3^{+}}{\Q_2\Q_3^-}\;\;,\;\;
\Y_2=H_R
\frac{\Q^{+}_3}{\Q_3^-}F_4^-\;,\nn\\
\Y_{4}=H_L \frac{1}{F_0^-} \frac{\Q_7^{+}}{\Q_7^-}\;\;,\;\;
\X_{4}=H_L\frac{\Q_7^{+}\Q_6^{--}}{\Q_7^- \Q_6 } \;\;,\;\;
\X_{3}=H_L\frac{\Q_6^{++}\Q_5^{-}}{\Q_6\Q_5^+}\;\;,\;\;
\Y_{3}=H_L
\frac{\Q^{-}_5}{\Q_5^+}\frac{1}{F_4^+}\;
\end{eqnarray}
\beq
{\rm where}\quad F_4=\prod_j\frac{x-x_{4,j}^+}{x-x_{4,j}^-}\;\;,\;\;
H_R=\lb\frac{x^{-}}{x^{+}}\rb^{\tfrac{L}{2}}
\prod_j\frac{x^+-x_{4,j}^-}{x^--x_{4,j}^-}
\sigma(u,x^{\pm}_{4,j})
\eeq
\beq
F_0=\bar F_4
\;\;,\;\;H_L=\tilde H_R\,.
\eeq
As before, the bar means the complex conjugation in  mirror plane
whereas the tilde is the complex conjugation in  physical plane.
The ${\cal Q}_a$ functions generalize the Baxter polynomials
- they are generic "polynomials" on the two-sheet Riemann surface\footnote{We allow for some of the Bethe roots $y_j$ to be at infinity.}
\beq\la{genpo}
{\cal Q}_a=\prod_{j=1}^{K_a} (x(u)-y_{a,j})\prod_{j=1}^{\bar K_a} \lb\frac{1}{x(u)}-y_{\bar a,j}\rb\;.
\eeq

The roots of these polynomials are constrained by the mirror reality condition of the
$Y$ functions. Namely, we have, as in the
strong coupling limit \eq{Xreality}\footnote{ and similarly for the left wing: \( \overline{\tilde{{\cal X}_{4}}}=\tilde{{\cal X}_{3}}\;\;,\;\;\overline{ \tilde{{\cal Y}_{4}}}
=\tilde{{{\cal Y}_{3}}}\)}
\beq
\overline{{\cal X}_{1}}={\cal X}_{2}\;\;,\;\;\overline{ {\cal Y}_{1}}
={\cal Y}_{2}\;.
\label{QXreality}
\eeq
This asymptotic solution has a few important symmetries and  analytic properties.
We will study some of them below. It is very important to find
a minimal set of such properties, such as reality and
analyticity, which can be used then to constrain
the $8$ functions parameterizing  the general solution, to generate only
the physically relevant solutions. We present below a possible
list of some of  such properties which, in our opinion, should be satisfied
by the physical solutions and try to constrain by them the  ABA solution  \eqref{eq:Q-monom}. This program worked well for the principal chiral field model \cite{Gromov:2008gj,Kazakov:2010kf} but it appears to be more tricky to do it for the AdS/CFT Y-system.

We will show that an essential part of  ABA can be derived from these properties.

\subsection{Minimal analyticity structure of Y-functions}
Here we summarize some of the analyticity properties of
Y-functions which, by our assumption, are satisfied by the physical solutions
of the AdS/CFT Y-system:\\
{\it\ Reality:}
\begin{enumerate}[I)]
\item Reality of Y-functions  $\bar Y_{a,s}=Y_{a,s}$
\item Reality of the Bethe roots \(u_{4,j}\)\footnote{We believe that the auxiliary roots should  be real or  appear in complex conjugated pairs,
though this question deserves a better study.}
\end{enumerate}
{\it\ Analyticity:}
\begin{enumerate}[1)]
\item $Y_{1,\pm 1},\;Y_{2,\pm 2}$ should have a Zhukovski cut on the real axes and be related by \eq{eq:Y11Y22}
\item $Y_{1,s}$ should have no branch cuts inside the strip $-\tfrac{s-1}{2}<\IM u<\tfrac{s-1}{2}$
\item $Y_{a,1}$ should have no branch cuts inside the strip $-\tfrac{a-1}{2}<\IM u<\tfrac{a-1}{2}$
\item $Y_{a,0}$ should have no branch cuts inside the strip $-\tfrac{a}{2}<\IM u<\tfrac{a}{2}$
\end{enumerate}
This list may  be not enough to completely constrain the ABA and the physical meaning of some of them remains to be understood. All this  deserves an additional study.
But these properties  are consistent with the
TBA equations for the excited states.

In what follows we consider for simplicity the operators/states obeying the symmetry $Y_{a,s}=Y_{a,-s}$.
The generalization to the full asymmetric case is almost straightforward.
One can see that $Y_{a,s}=Y_{a,-s}$ implies (which is also true for finite $L$)
\beq
\frac{\X_4^{+}}{\Y_4^{+}}=\frac{\Y_1^-}{\X_1^-}\;\;,\;\;\frac{\X_4}{\X_3}=\frac{\X_2}{\X_1}
\;\;,\;\;\frac{\X_3^{-}}{\Y_3^{-}}=\frac{\Y_2^+}{\X_2^+}\;.
\eeq

 \subsubsection{Reality}\la{sec:re}
 Reality of $Y_{a,s>0}$ implies that $T_{a,s}$ are also real up to a gauge transformation.
 Here we will examine this condition
 in the asymptotic large \(L\) limit.

It is easy to see that since the first four
functions $\X_1,\Y_1,\Y_2,\X_2$
are small
whereas $\X_3,\Y_3,\Y_4,\X_4$ are large
in the $L\to\infty$ limit
only a half of
the generating functional \eq{eq:GenFun224} (corresponding to one
of the subgroups
$su(2|2)_{L,R}$) contributes.
For $s\ge 0$ the full functional reduces to \cite{Beisert:2006qh}
\footnote{For $s\le 0$ only the complimentary part  of the full generating functional, dropped in \eqref{eq:WR}, is relevant.  }
\beq\label{eq:WR0}
{\cal W}^R\simeq (1-D\Y_1 D)
\frac{1}{1-D\X_1 D}
\frac{1}{1-D\X_2 D}
(1-D\Y_2D)\frac{\Y_3^-\Y_4^{+}}{\X_3^-\X_4^{+}}
=\sum_{s=0}^{\infty} D^{s}T_{1,s}D^{s}
\eeq
Since we fixed $T_{0,s}=1$ we have only two degrees of freedom left:
one is a possible redefinition of $D\to g(u) D$
which does not change the definition of the shift operator, whereas another corresponds to the  transformation is
${\cal W}\to {\cal W} g(u)$.
In particular, we can remove the last factor from \eqref{eq:WR0}
by a desired gauge transformation \eq{eq:gauge} to get
\beq\label{eq:WR}
{\cal W}=(1-D\Y_1 D)
\frac{1}{1-D\X_1 D}
\frac{1}{1-D\X_2 D}
(1-D\Y_2D)\;.
\eeq

Let us show that the hermiticity of the above generating functional
automatically implies the reality of all $Y_{a,s> 0}$. Indeed since
\(D\) is hermitian we have
\beqa
{\cal W}&=&\sum_s D^s T_{1,s} D^s=\sum_sD^s \bar T_{1,s} D^s={\cal W}^\dagger\\
&=&(1-D\bar\Y_2 D)
\frac{1}{1-D\bar\X_2 D}
\frac{1}{1-D\bar\X_1 D}
(1-D\bar\Y_1D)\;,\nn
\eeqa
which is equivalent to \eq{QXreality}:
since we have to equate the coefficients of infinitely many powers of $D$ the monomials should  coincide.
This relation is a quantum analog of \eq{eq:Z4}. Notice that
\eq{QXreality} implies (assuming that $H_R$ and $F_4$ do not depend explicitly
on the Bethe roots $y_{a,j},y_{\bar a,j}$) that
\beq\la{recon}
\bar \Q_1= \Q_3\;,\quad\bar \Q_2= \Q_2
\;,\quad {\bar F_4}={F_0}\;,\quad
{\bar H_R}={H_R}\;.
\eeq
The first equality implies that $y_{\bar 1,j}=y_{3,j},\;y_{1,j}=y_{\bar 3,j}$. The second
equality tells us that $y_{2,j}=y_{\bar 2,j}$ i.e. that $\Q_2$ is a usual polynomial
of $u$. Notice that the combination $\Q_1 \Q_3$ is a polynomial of $u$.
Finally, the last equality implies for the (unitary) dressing factor
\beq\la{crs}
\prod_j
\frac{\sigma(u,u_{4,j})}{\bar \sigma(u,u_{4,j})}=
\prod_j\frac{1/x^--x_{4,j}^+}{1/x^+-x_{4,j}^+}
\frac{x^--x_{4,j}^-}{x^+-x_{4,j}^-}\;
\eeq
which is the crossing condition of \cite{Janik:2006dc}!\footnote{The original
crossing relation of Janik coincides with \eq{crs} up to a factor which becomes \(1\) due to the zero total momentum (level matching) condition on the roots \(u_{4,j}\).}
(see     \cite{RewIII3} for its solution).

\subsubsection{Analyticity properties 1), 2)}\label{ssec:anal}
To see the consequences of the analyticity property 1) let us make a simple observation.
By a direct calculation of the corresponding T-functions from \eqref{eq:WR}
we get
\beq\la{F0F4}
Y_{1,+1}Y_{2,+2}\simeq\frac{\X_1^-\X_2^+}{\Y_1^-\Y_2^+}=\frac{1}{F_0 F_4}\;,
\eeq
 and hence  the property 1) immediately implies
\(
F_0(u+i0) F_4(u+i0)=\frac{1}{F_0(u-i0) F_4(u-i0)}\;.
\)
To arrive to the above conclusion we used a weaker version of the property 1)
for the product of two $Y$ functions. In fact, one can get more from the property 1), namely
\(F_4^{[+0]}=1/F_0^{[-0]}\;
\),
which together with \eq{recon} gives a powerful constraint on the functions
$F_4$ and $F_0$. We will show below that requiring $F_4$ and $F_0$
to have only one Zhukovski cut on the real axes leads to the conditions 2) and 3).

Let us study the property 2). Notice that the transformation $\X_{1,2}\to g\X_{1,2},\;\Y_{1,2}\to \;g\Y_{1,2}$
where \(g(u)\) is an arbitrary function, does not affect  $Y_{1,s}$ and therefore it is a gauge transformation. If we take $g=\frac{\Q^+_1}{H_R\Q_1^-}$ we notice that
$\Q_1$ appears only in the combination $\Q_1 \Q_3$, which
does not have branch cuts  as we have shown above.
This implies that in that gauge $\X_1$ and $\X_2$ have no branch cuts any more.
Expanding the denominator in \eq{eq:WR} we get\\
\(
W\simeq\sum\limits_{s=0}^\infty D^s(1-D\Y_1^{[+s]} D)
\lb \sum_{n=-s}^s \X_1^{[-s+1]}\dots  \X_1^{[n-1]}\X_2^{[n+1]}\dots\X_2^{[+s-1]} \rb
(1-D\Y_2^{[-s]}D)D^s.\nn
\)
   The cuts in $T_{1,s}$ come only from  $\Y_1^{[+s-1]}$
and $\Y_2^{[-s+1]}$.
The analyticity requirement  2)
  is satisfied since $T_{1,s}\sim \Y_1^{[+s-1]}\Y_2^{[-s+1]}\sim F_0^{[+s]}F_4^{[-s]}$
have the analyticity strip $|\IM(u)|<s/2$
because $F_0,F_4$ have only a single cut
one the real axes.

\subsubsection{Duality transformation and analyticity 3)}
  Similarly to the
Sec.\ref{sec:du} we can consider  a duality transformation as an effect of the commutation
of two operatorial factors within the generating functional:
\beq\label{eq:DUAL}
(1-D\X_1 D)^{-1}
(1-D\Y_1 D)=
(1-D\hat\Y_1 D)
(1-D\hat\X_1 D)^{-1}
\eeq
and a similar equation for the factors with $\X_2,\Y_2$.
It is convenient to parameterize the new factors as (compare it with \eqref{eq:paramQ},\eqref{eq:DualSU22})\
\beq\la{eq:afterd}
\hat\X_1=\hat H_R\frac{\hat\Q_1^+}{\hat\Q_1^-}\frac{1}{F_0^-}\;\;,\;\;
\hat\Y_1=\hat H_R\frac{\Q_2^{--}\hat\Q_1^+}{\Q_2\hat\Q_1^-}\;\;,\;\;
\hat\Y_2=\hat H_R\frac{\Q_2^{++}\hat\Q_3^-}{\Q_2\hat\Q_3^+}\;\;,\;\;
\hat\X_2=\hat H_R\frac{\hat\Q_3^-}{\hat\Q_3^+}\frac{1}{F_4^+}
\eeq
where by definition $\hat H_R$ can only depend on the momentum carrying   roots $u_{4,k}$
whereas $\hat Q_1$ is a function of the form \eq{genpo}.
In this parameterization we have
$\hat\X_1^+/\hat\Y_1^+=\X_1^-/\Y_1^-,\;\hat\X_2^-/\hat\Y_2^-=\X_2^+/\Y_2^+$
and to keep   \eq{eq:WR} intact we only have  to satisfy commpare with \eqref{eq:XYdual})
\(
\hat\X_1-\hat\Y_1=\X_1-\Y_1\;\;,\;\;
\hat\X_2-\hat\Y_2=\X_2-\Y_2
\)
Second equation is the complex conjugate of the first one.  The first equation gives
\(
\frac{\hat\Q_1^+\Q_1^+}{\hat\Q_1^-\Q_1^-}=
F_0^-\frac{H_R}{\hat H_R}\,\,\,
\frac{
F_0^+\Q_2-{\Q_2^{++}}
}{
F_0^-\Q_2^{--}
-\Q_2}
\)
which has the solution
\(
\hat\Q_1\Q_1=f(u,x_{4,k})(F_0\Q_2^--\Q_2^+)\,.
\)
Since the r.h.s. has no poles at $x=1/x_{4,k}^+$ and cannot explicitly depend on $x_{4,k}$
we should take $f=C\,\prod_{k}(1/x-x_{4,k}^+)$ to cancel the poles in $F_0$. This leads to the condition
\(
\hat \Q_1\Q_1\propto
\Q_2^+\prod_{k=1}^{K_4}(1/x-x_{4,k}^+)-{\Q_2^{-}}\prod_{k=1}^{K_4}(1/x-x_{4,k}^-)
\)
from where we can determine $\hat\Q_1$ and its complex conjugate $\hat\Q_3$.
The resulting formulas are   analogous to \eq{eqQQ}.

We demonstrated above that the terms in the generating functional can be
reshuffled in such a way that the expression for the new elementary factors \eq{eq:afterd}
are very similar to the initial ones \eq{eq:Q-monom}, with the modified Bethe roots.

Now let us use this fact to show that
$Y_{a,1}$ are also analytic in their strips given in the property 3). Indeed, using the B\"acklund relations, in the way similar to the subsection \ref{ssec:GeNFun} where \(T_{a,s}\) was generated from
\({\cal W},\)    we can show that
  $T_{a,1}$ can be computed from
${\cal W}^{-1}$  as follows \cite{Kazakov:2007fy}
\beq\label{eq:Wminus1}
{\cal W}^{-1}=(1-D\hat\X_2 D)
\frac{1}{1-D\hat\Y_2 D}
\frac{1}{1-D\hat\Y_1 D}
(1-D\hat\X_1D)\simeq\sum_{a=0}^{\infty} (-1)^aD^{a}T_{a,1}D^{a}\;.
\eeq
As we saw in subsection \ref{ssec:anal},  for the analyticity of T-functions in their physical strips   $F_4$ should have a cut  only on the real axes. The
arguments given there can be also applied to the functional \eqref{eq:Wminus1} which leads to the proof of the property  3).
It also shows  that in a certain   gauge $T_{a,1}$ has the
analyticity strip $|\IM(u)|<a/2$.

Due to \eqref{eq:wrapY}  we can drop the denominator in the r.h.s. of
\eq{eq:Ysystem} at \(s=0\) and rewrite it,
  using \(1+Y_{a,s}=\frac{T_{a,s}^+T_{a,s}^-}{T_{a+1,s}T_{a-1,s}}\) following
from \eqref{eq:HirotaKM} and \eqref{Ydef},
$
\frac{Y_{a,0}^+Y_{a,0}^-}{Y_{a-1,0}Y_{a+1,0}}\simeq\left(\frac{T_{a,1}^+T_{a,1}^-}{T_{a-1,1}T_{a+1,1}}\right)^2\,,
$
where in the equation for $a=1$ one should replace in the l.h.s. $Y_{0,0}$ by $1.$ Solving this Y-system equation
for $Y_{a,0}$ we get
\(
Y_{a,0}=\frac{\phi(u+ia/2)}{\phi(u-ia/2)}T_{a,1}^2
\)
where the first factor,  a zero mode, is easy to calculate  since \(\phi(u)\) can be extracted from   $Y_{1,0}$. Hence  the most complicated part of $Y_{a,0}$
is hidden in $T_{a,1}$ and has the correct analyticity structure 4). The
proof of the correct analyticity  of the factor
 \(\frac{\phi(u+ia/2)}{\phi(u-ia/2)}\) is left to the reader.

\subsubsection{Reality of \texorpdfstring{$Y_{a,0}$}{Ya0}}

\comment{
\subsubsection{Left-right wing exchange symmetry}
We can repeat similar arguments for the left \(SU(2|2)_L\) wing in the general solution \eqref{eq:GenFun224}. This wing should be expanded in powers of \(D^{-1}\). By simple operatorial manipulations we rewrite it as
\beq\label{eq:WL}
W^L=\frac{\Y_3^-}{\X_3^-}\lb1-D^{-1}\frac{\X_3^{--}}{\Y_3^{--}\X_3}D^{-1}\rb
\frac{1}{1-D^{-1}\frac{1}{\X_3}D^{-1}}
\frac{1}{1-D^{-1}\frac{1}{\X_4}D^{-1}}
\lb1-D^{-1}\frac{\X_4^{++}}{\Y_4^{++}\X_4}D^{-1}\rb
\frac{\Y_4^+}{\X_4^+}
\eeq
Due to the fact that we limited ourselves by solutions obeying the  left-right wing exchange symmetry, occurring as the  \(s\leftrightarrow -s\) of the T-functions, the transposed generating functional for the left wing should
coincide up to a gauge transformation with the right wing generating functional
\beq
{(W^L)}^T=\frac{\Y_4^+}{\X_4^+}
\lb1-D\frac{\X_4^{++}}{\Y_4^{++}\X_4}D\rb
\frac{1}{1-D\frac{1}{\X_4}D}
\frac{1}{1-D\frac{1}{\X_3}D}
\lb1-D\frac{\X_3^{--}}{\Y_3^{--}\X_3}D\rb
\frac{\Y_3^-}{\X_3^-}
\eeq
i.e. comparing it with \eqref{eq:WR} we get the following relations
\beq
\frac{\X_4^{+}}{\Y_4^{+}}=\frac{\Y_1^-}{\X_1^-}\;\;,\;\;\frac{\X_4}{\X_3}=\frac{\X_2}{\X_1}
\;\;,\;\;\frac{\X_3^{-}}{\Y_3^{-}}=\frac{\Y_2^+}{\X_2^+}
\eeq
i.e.
\begin{eqnarray*}
\Y_{4}=H^L \frac{1}{F_0^-} \frac{\Q_1^{+}}{\Q_1^-}\;\;,\;\;
\X_{4}=H^L\frac{\Q_1^{+}\Q_2^{--}}{\Q_1^- \Q_2 } \;\;,\;\;
\X_{3}=H^L\frac{\Q_2^{++}\Q_3^{-}}{\Q_2\Q_3^+}\;\;,\;\;
\Y_{3}=H^L
\frac{\Q^{-}_3}{\Q_3^+}\frac{1}{F_4^+}
\end{eqnarray*}
where $H^L$ is some new function.
}
\comment{
Let us now see the consequences
of the reality of middle node Y-functions \(Y_{a,0}\). From the asymptotic generating functionals \eqref{eq:WR},\eqref{eq:WL} we have\beqa
T_{1,+1}&\simeq&\frac{\Y_3^{--}\Y_4}{\X_3^{--}\X_4}(-\Y_1+\X_1+\X_2-\Y_2)\;\;,\;\;
T_{1,+0}\simeq\frac{\Y_3^-\Y_4^{+}}{\X_3^-\X_4^{+}}\\
T_{1,-1}&\simeq&\frac{\Y^{--}_3\Y_4^{++}}{\X^{--}_3\X_4^{++}}\lb-\frac{\X_3^{--}}{\Y_3^{--}}\frac{1}{\X_3}+\frac{1}{\X_3}+\frac{1}{\X_4}-\frac{\X_4^{++}}{\Y_4^{++}}\frac{1}{\X_4}\rb
\eeqa
i.e.
\beq
Y_{1,0}=\frac{\X_3\Y_3^{--}}{\Y_3\X_3^{--}}
\lb-\frac{\X_3^{--}}{\Y_3^{--}}\frac{1}{\X_3}+\frac{1}{\X_3}+\frac{1}{\X_4}-\frac{\X_4^{++}}{\Y_4^{++}}\frac{1}{\X_4}\rb
(-\Y_1+\X_1+\X_2-\Y_2)^2
\eeq
or}
Finally let us also imply the reality condition to
\(
Y_{1,0}=\frac{(\Y_1-\X_1-\X_2+\Y_2)^2}{\Y_2\Y_3}
\).
Note that the numerator is real,  so for the reality of \(Y_{1,0}\) we only  have to require that
\(
\Y_2\Y_3=H_L H_R\frac{F_4^-}{F_4^+}
\)
is real.
Note that the factors $H_R$ is a real function as a consequence of the crossing equation  in the mirror kinematics stemming from the last of eqs.\eq{recon}. In the physical kinematics $H_L$
is conjugate to $H_R$ and naively one would expect
it to be also real as well. However the conjugation in the physical sense does not
necessarily commute with the mirror conjugation.
Explicit calculation shows that
\(
\Y_2\Y_3=\frac{Q_4^{++}Q_4^{--}}{Q_4^2}\;\;,\;\;Q_4=\prod_{k=1}^{K_4}(u-u_{4,k})
\)
which is indeed real. The reality of all  other $Y_{a,0}$ follows from the Y-system.

Finally, the above expression for \(Y_{1,0}\) simplifies on a Bethe root \(u=u_{4,k}\):  since \(1/F_4^-(u_{4,k})=0~\),
\(\Y_2\)   dominates the numerator and we get from
\(Y_{1,0}(u_{4,k})=\Y_2/\Y_3=\frac{H_R}{H_L}F_4^+F_4^-\frac{\Q_3^+\Q_5^+}{\Q_3^-\Q_5^-}=-1\).
Since $H_R$ is a complex conjugate of $H_L$ we see that  $Y_{1,0}(u_{4,k})$ is unimodular
in the physical kinematics, ensuring the reality of the Bethe roots $u_{4,k}$\footnote{This property was also demonstrated in \cite{Gromov:2009zb}  numerically to hold
 at finite $L$. }.

\section{Conclusions and perspectives}

Our main purpose in these notes was mainly pedagogical: to show the power of Hirota discrete integrable dynamics (HDID) for the solution of quantum integrable models.  The B\"acklund method of solution of Hirota equation for fusion in the supersymmetric generalizations of the Heisenberg spin chain, with
the polynomiality condition of the transfer matrix
 gives a rather direct way of derivation of the final
 Bethe ansatz equations for the roots of Baxter's Q-polynomials. However, the applications of HDID is not limited to the spin chains and can be quite efficient in the study of integrable CFT's,  \(\s\)-models at finite
 volume and, remarkably, in such a complicated problem as the AdS\(_5\)/CFT\({_4}\) system.

 We demonstrated that the
 general asymptotic solution of Y-system for AdS/CFT obeys several remarkable analyticity and reality
 properties. They seem to be rather constraining and could be
 used at finite volume to single out the physically relevant solutions of this Y-system. It seems possible to reverse the logic
 and derive the asymptotic Bethe ansatz equations  from these equations,
  as  in  relativistic \(\s\)-models.

    We hope that  further study of this circle of questions, and in particular
    of the role and the consequences of  the  equation \eqref{eq:Y11Y22}, will lead to the complete understanding of the analyticity structure of the
    integrable AdS/CFT systems. This understanding, together with the solutions of Y-system stemming from the HDID (in the form the generating functional \eqref{eq:GenFun224}  or in the Wronskian form recently obtained in \cite{Gromov:2010km}) should allow to reduce the problem to a
    finite set of integral Destri-DeVega type equations.

\vskip0.5cm

{\bf Acknowledgments}

 The work  of VK was partly supported by  the ANR grants INT-AdS/CFT\\ (ANR36ADSCSTZ)  and   GranMA (BLAN-08-1-313695) and the grant RFFI 08-02-00287.  V.K. also thanks NORDITA institute in Stokholm, as well as  Vladimir Bazhanov and  the theoretical physics group of  Australian National University (Canberra) for the kind hospitality and interesting discussions.  We  thank Sebastien Leurent, Zengo Tsuboi,
 Pedro Vieira and especially Dymitro Volin for useful  discussions.

%\bibliography{chapters,template}
%\bibliographystyle{nb}

%\bibliography{intads}
%\bibliographystyle{nb}

%%%%%%%%%%%%%%%%%%%%%%%%%%%%%%%%%%%%%%%%
\phantomsection

\end{document}